\documentclass[runningheads]{llncs}
\usepackage[T1]{fontenc}
\usepackage{graphicx}
\usepackage{xcolor}
\usepackage{subfig}
\begin{document}
\title{Enhancing Representation Learning of EEG Data with Masked Autoencoders}

\titlerunning{Masked Autoencoders for EEG data}

\author{Yifei Zhou,
Sitong Liu}
\authorrunning{Y. Zhou, S. Liu}
%
\institute{George Washington University, Washington, DC 20052, USA \\
\email{\{yzhou87,sitong.liu\}@gwmail.gwu.edu}}
\maketitle              
\begin{abstract}
Self-supervised learning has been a powerful training paradigm to facilitate representation learning. In this study, we design a masked autoencoder (MAE) to guide deep learning models to learn electroencephalography (EEG) signal representation. Our MAE includes an encoder and a decoder. A certain proportion of input EEG signals are randomly masked and sent to our MAE. The goal is to recover these masked signals. After this self-supervised pre-training, the encoder is fine-tuned on downstream tasks. We evaluate our MAE on EEGEyeNet gaze estimation task. We find that the MAE is an effective brain signal learner. It also significantly improves learning efficiency. Compared to the model without MAE pre-training, the pre-trained one achieves equal performance with 1/3 the time of training and outperforms it in half the training time. Our study shows that self-supervised learning is a promising research direction for EEG-based applications as other fields (natural language processing, computer vision, robotics, etc.), and thus we expect foundation models to be successful in EEG domain.

\end{abstract}
\section{Introduction}

Electroencephalography (EEG) data, with its rich multidimensional structure, offers unique insights into various neurological phenomena~\cite{murungi2023empowering}. Understanding the complexities of human brain activity through EEG signals has long been a focal point in neuroscience. EEG-based research holds immense potential for decoding cognitive processes, mental states, and various spatial and temporal aspects of brain functioning.

EEG is widely utilized in research areas such as neural engineering, neuroscience, biomedical engineering, and brain-like computing, particularly in applications like brain-computer interfaces (BCIs). The analysis of EEG signals is fundamental to the advancement of BCIs, offering profound insights into the intricate neural processes of the human brain. Over the past decade, numerous machine learning and deep learning algorithms have been employed to analyze EEG data, resulting in significant advancements in various applications. These include emotion recognition, motor imagery, mental workload assessment, seizure detection, Alzheimer's classification, sleep stage scoring, and many others~\cite{craik2019deep,kastrati2021eegeyenet,roy2019deep,altaheri2023deep,qu2022time,gao2021complex,hossain2023status,yi2022attention,key2024advancing,li2024enhancing,koome2023trends,zhou2022brainactivity1,qu2020identifying,qu2020using,qu2020multi,qu2018eeg,qu2019personalized,saeidi2021neural,qu2022eeg4home,rasheed2020machine,dadebayev2022eeg,wang2022eeg,li2020deep,aggarwal2022review}.

EEG and deep learning research have been in close proximity for decades, with advancements in both fields contributing to strides in our understanding of the brain. Deep learning algorithms are particularly popular in the context of EEG analysis due to their ability to extrapolate and generalize input information, making them ideal for decoding the complexities and noise within EEG signals into interpretable outputs.

The EEGEyeNet dataset~\cite{kastrati2021eegeyenet} has become a cornerstone in the realm of cognitive neuroscience and machine learning, facilitating the advancement of eye-tracking technologies through the integration of EEG data. The fusion of these disciplines aims to enhance the precision of eye position prediction, a critical aspect in understanding visual attention and neurological behavior. However, the integrity of the EEGEyeNet dataset is compromised by the presence of data points exhibiting eye positions that surpass the physical boundaries of the experimental screen, leading to potential inaccuracies in subsequent analyses and model training.

Among the numerous EEG-based tasks, gaze position estimation is a significant challenge due to its relevance in spatial cognition. This task is performed based on the \textit{Large Grid Paradigm} where participants are instructed to focus on a succession of dots that appear one after another, with each dot appearing at one of 25 distinct positions on the screen~\cite{kastrati2021eegeyenet}. The task is to predict the XY-coordinate of the participant's gaze position. Accurate decoding of absolute positions from EEG signals holds implications for neurorehabilitation, brain-computer interfaces, and understanding fundamental aspects of spatial awareness.

\begin{figure}[t]
    \centering
    \subfloat[Pre-training model]{
        \includegraphics[width=\textwidth]{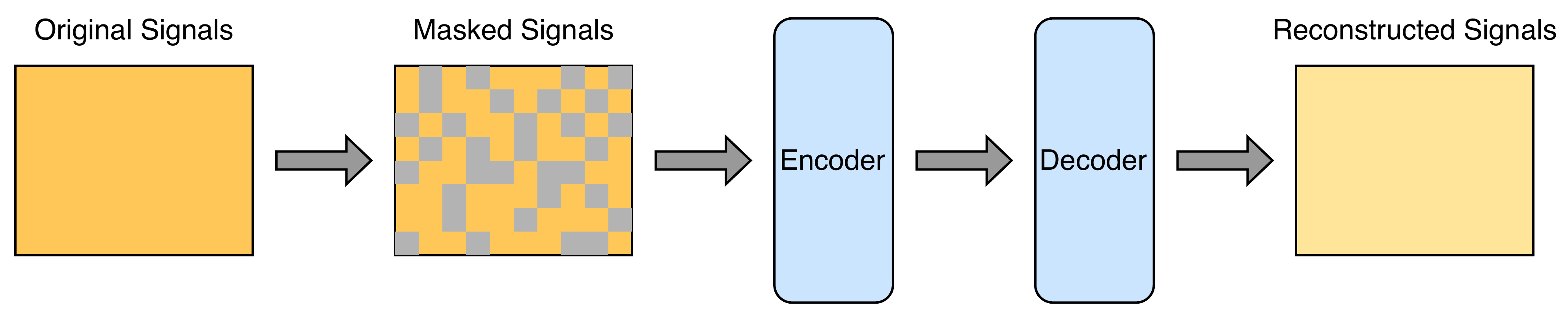}
        \label{fig:pre-train}
    }
    \vspace{5mm}
    \subfloat[Fine-tuning model]{
        \includegraphics[width=0.5\textwidth]{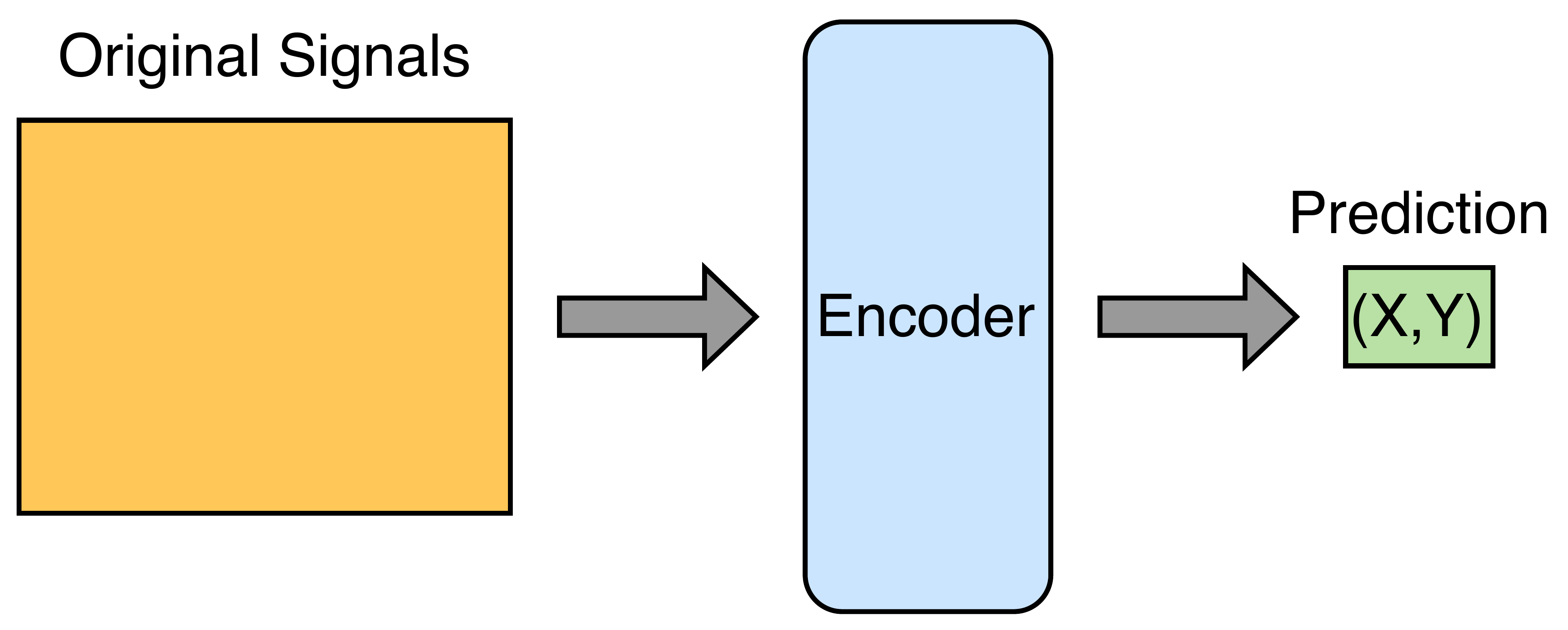}
        \label{fig:fine-tune}
    }
    \caption{\textbf{Pre-training and fine-tuning model architectures.} EEG signals collected from multiple channels are arranged into a matrix. \textbf{(a)} We mask random elements from the input EEG signal matrix. Our MAE learns to recover these missing signals. \textbf{(b)} Our main purpose is to measure the encoder's performance change after MAE pre-training, so we remove the decoder and fine-tune the encoder to predict gaze positions.}
    \label{fig:architecture}
\end{figure}

Deep learning methodologies have shown remarkable promise in unraveling intricate patterns within EEG data \cite{roy2019deep,altaheri2023deep}. Recently, the widely-used Vision Transformer (ViT) model~\cite{dosovitskiy2020image} has been proven to be able to significantly improve the accuracy of absolute position prediction~\cite{yang2023vit2eeg}. The model proposed by this study, EEGViT, provides further evidence that EEG-based tasks could benefit from computer vision models. EEGViT leverages ViT model weights pre-trained on the ImageNet dataset~\cite{deng2009imagenet} to achieve state-of-the-art performance, demonstrating that pre-training can contribute to the success of the model in addition to the model architecture~\cite{yang2023vit2eeg}. Our study further explores the potential of pre-training to boost the model performance without data augmentation or modifying the model architecture.

Self-supervised pre-training is a prevailing practice to facilitate the representation learning of deep learning models. It helps the models learn useful patterns and representation from the data and thus the models achieve better performance on downstream tasks. In natural language processing (NLP), self-supervised pre-training has been employed to guide large language models to learn contextual information from text corpora \cite{radford2018improving,radford2019language,kenton2019bert,brown2020language,openai2023gpt}. Inspired by BERT~\cite{kenton2019bert}, masked autoencoder (MAE) is applied to computer vision models and shown to be successful and scalable vision learners \cite{dosovitskiy2020image,bao2021beit,he2022masked}.

As a self-supervised pre-training technique, MAE removes certain ratios of content from inputs and tries to reconstruct them. When it is applied in ViT, a certain ratio of input image patches are masked, and the goal is to recover these masked patches~\cite{he2022masked}. Since EEGViT has shown the capability of the ViT on EEG data, the applicability of MAE on EEG data is worth studying as well. Therefore, our research question is: are MAEs effective brain signal learners? We attempt to answer this question by employing a MAE design that is similar to the one used for ViT pre-training. Our MAE masks random signals from the input EEG signal matrix and reconstructs the missing signals. It has an encoder-decoder architecture (Figure \ref{fig:pre-train}). The encoder operates on masked EEG signals and learns meaningful latent representations. The decoder then reconstructs the input signals from these latent representations. After pre-training with our MAE, the decoder is removed and the encoder is applied to unmasked EEG signals for gaze position prediction (Figure \ref{fig:fine-tune}).

We compare the performance of the encoder pre-trained with our MAE to the encoder trained from scratch. Experiment results show that MAE pre-training boosts the encoder's performance on EEGEyeNet gaze estimation task without data augmentation or modifying the encoder architecture. Compared to the encoder without MAE pre-training, the pre-trained one achieves equal performance with 1/3 the time of training and outperforms it in half the training time (Figure \ref{fig:efficiency}). We anticipate that EEG-based applications will benefit more from self-supervised pre-trained deep learning models just as other fields (NLP, computer vision, robotics, etc.), and this even suggests the promising research on foundation models \cite{bommasani2021opportunities,yang2023foundation,zhou2023comprehensive,li2023multimodal,firoozi2023foundation} in the EEG domain.

\section{Related Work}
\subsection{Masked modeling in language and vision}
Self-supervised pre-training by masked modeling has brought huge progress to natural language processing (NLP). The masking mechanism in BERT~\cite{kenton2019bert} is to randomly mask a certain percentage of the input tokens, and train the model to predict the original token that has been masked out. GPT \cite{radford2018improving,radford2019language,brown2020language,openai2023gpt} adopts an autoregressive training approach that predicts the next word in a sentence given all the previous words, which means that during training, the model looks at a part of a sentence and learns to predict the word that comes next. Inspired by the practices in NLP, masked encoding has been applied to visual representation learning \cite{chen2020generative,dosovitskiy2020image,bao2021beit,he2022masked}.

\subsection{Masked autoencoder for EEG data}
Various deep learning models such as convolutional neural network (CNN), recurrent neural network (RNN) and Transformer have been applied to EEG data \cite{bashivan2015learning,craik2019deep,roy2019deep,mao2020eeg,kostas2021bendr,yi2022attention,xiao20224d,weng2023interpretable,yang2023vit2eeg,abibullaev2023deep,xie2022transformer,sun2021eeg,gong2023eeg,key2024advancing,li2024enhancing,koome2023trends,murungi2023empowering,qu2022eeg4home,dou2022time,zhou2022brainactivity1,wang2022eeg,qu2022time,qu2020identifying,qu2020using,qu2020multi,qu2018eeg,qu2019personalized}. While supervised learning has been a dominant paradigm of training large deep learning models for a decade, in recent years, self-supervised pre-training by masked modeling has been a great performance booster. A deep learning model pre-trained with masked autoencoders (MAE) often outperforms the same model solely trained with supervised learning. The success of MAE in NLP and computer vision suggests that it is an effective representation learner for both temporal and spatial data. Therefore, it is a natural idea to apply MAE to EEG data.

Previous work has demonstrated the advantage of MAE on EEG-based sleep stage classification~\cite{chien2022maeeg}, seizure sub-type classification~\cite{peng2023wavelet2vec} and cognitive load classification~\cite{pulver2023eeg}. The MAEs in these studies reconstruct original features or raw signals from masked \textit{features}. Our study, however, employs a simple approach that reconstructs original EEG signals from masked \textit{signals}. The input EEG signals are directly masked and fed to our MAE without further preprocessing and feature extraction. Experiments have shown that this simple design can still guide our MAE to learn signal representation that is useful for downstream tasks.

\subsection{EEG-based gaze estimation}
\label{task}
EEG-based gaze estimation aims at combining EEG signals with computational techniques to predict the direction or position of a person's gaze. This approach leverages the fact that certain patterns in brain activity, as captured by EEG, correlate with where a person is looking. 

The EEGEyeNet dataset~\cite{kastrati2021eegeyenet} is a comprehensive collection of high-density, 128-channel EEG data synchronized with eye-tracking recordings from 356 healthy adults. This dataset is unique due to its large scale and precise annotation, encompassing over 47 hours of recording. The third task in the associated benchmark involves determining the absolute position of the subject's gaze on a screen, described in terms of XY-coordinates. This task is performed using data from the Large Grid paradigm, where participants fixate on a series of dots at different screen positions. It is the most challenging task in the benchmark, aiming to simulate a purely EEG-based eye-tracker. The performance is measured as the euclidean distance in millimeters between the actual and the estimated gaze position. Current performance of deep learning models on this task is presented in Table 4 of~\cite{yang2023vit2eeg}.

\section{Methods}

We design a masked autoencoder (MAE) that randomly masks signals from the input EEG signal matrix and recovers these missing signals. As shown in Figure \ref{fig:architecture}, our MAE has an encoder-decoder architecture. The encoder operates on masked EEG signals and learns meaningful latent representations. The decoder then reconstructs the input signals from these latent representations. As the overall goal is to enhance the encoder's capability to learn useful signal representations, after MAE pre-training, the decoder is removed and the encoder is applied to unmasked EEG signals to perform downstream tasks. By doing so, we are able to measure the encoder's performance change after MAE pre-training.

\subsection{Masking mechanism}
\label{masking}
The masking is applied based on the matrix representation of EEG signals. Raw EEG signals are collected from multiple channels. The signals from each channel can be stacked row by row to form a matrix that is suitable for being neural network input~\cite{yang2023vit2eeg}. 

Before an EEG signal matrix is sent to our MAE encoder, a certain proportion of its elements are randomly selected to be set to zero. We implement a simple random selection. Suppose the dimension of EEG signal matrices is $m\times n$ and the masking ratio is $r$. First we generate a random permutation of integers from 0 to $m\times n-1$. Then we select the first $m\times n\times r$ integers from this permutation as the indices to be masked. Next these selected indices are converted into 2D indices corresponding to the row and column dimensions of the EEG signal matrix. For index $i$ in the selected indices, its corresponding row index is $\left\lfloor \frac{i}{n} \right\rfloor$ and column index is $i\bmod n$. The corresponding elements in the EEG signal matrix will be set to 0. 

During training, a mask is generated for each batch and epoch, which means that none of the previously used masks is directly reapplied to the current batch. This will avoid overfitting by ensuring that our MAE can learn as rich local and global patterns as possible. The MAE cannot solve the reconstruction task by simply memorizing the signal values.

\subsection{Encoder design}
Our MAE encoder is EEGViT~\cite{yang2023vit2eeg}, a hybrid Vision Transformer (ViT) architecture designed for EEG data. It combines a two-step convolution block~\cite{lawhern2018eegnet} with the ViT layers. When the ViT layers are initialized with the model weights pre-trained on ImageNet dataset~\cite{deng2009imagenet}, EEGViT achieves state-of-the-art performance (Table 4 of~\cite{yang2023vit2eeg}).

The visual knowledge that ViT learns from large image datasets is beneficial to EEG data as well. However, EEGViT utilizes pre-trained ViT model weights directly for supervised training. We believe that the ViT model can first learn some general EEG signal knowledge before it is applied to a specific task at hand, by which the model can experience a milder transfer from vision domain to EEG. We bridge this gap by using pre-trained ViT weights for MAE pre-training. The ViT layers in our encoder are initialized with the model weights pre-trained on ImageNet dataset. After the encoder learns general EEG signal representation, it will be fine-tuned on downstream tasks.

\subsection{Decoder design}
\label{decoder}
Following the MAE for ViT~\cite{he2022masked}, our MAE decoder is a series of Transformer blocks. The reason for this choice resembles the one for vision MAE. Our reconstruction task is at \textit{signal} level. It requires a low-level understanding of EEG raw signals. A low-level reconstruction task like pixels, or in our case, signals, needs a non-trivial decoder architecture. As described in~\cite{he2022masked}, the decoder design determines the semantic level of learned information. Different decoder structures drive the encoder to extract different levels of signal patterns.

As introduced before, in the fine-tuning stage, only the encoder is kept for supervised training. The MAE decoder assists the encoder with efficient signal encoding, but since our main purpose is to compare the encoder's performance before and after MAE pre-training, the decoder is not used for downstream tasks.

\subsection{Reconstruction task}
Our MAE takes in masked EEG raw signals and outputs reconstructed signals. Note that we aim to recover the missing signals, but for implementation simplicity the unmasked signals are also "reconstructed". That is, our MAE output has the same dimension as the input. Since we only care about the recovery of missing signals, the reconstruction loss is computed on the masked elements of an EEG signal matrix. This practice is similar to previous work \cite{kenton2019bert,he2022masked}.

Following MAEEG~\cite{chien2022maeeg}, we adopt a similarity loss function\footnote{We also experiment with mean squared error (MSE) loss function, the performance increase brought by it is not obvious.}:
\begin{equation}
    \mathcal{L}=1-\frac{\hat{\mathbf{x}} \cdot \mathbf{x}}{\|\hat{\mathbf{x}}\|\|\mathbf{x}\|}
\end{equation}
where $\mathbf{x}$ is the original signals and $\hat{\mathbf{x}}$ is the reconstructed signals. $\frac{\hat{\mathbf{x}} \cdot \mathbf{x}}{\|\hat{\mathbf{x}}\|\|\mathbf{x}\|}$ computes cosine similarity. Subtracting it from 1 ensures that our MAE learns to minimize the reconstruction loss. Cosine similarity encourages our MAE to capture the intrinsic characteristics of EEG signals. We apply a reversed mask to both the MAE output and full input, so that previously masked positions are now retained and unmasked positions are now set to zero. Then we flatten these two matrices to compute the loss.

\section{Experiment Setting}
We use the EEGEyeNet dataset~\cite{kastrati2021eegeyenet} for MAE pre-training. Then we fine-tune all layers of the MAE encoder on the same dataset.

\subsection{EEG data}
The EEG data for training our model are from "Large Grid Paradigm" in EEGEyeNet dataset which involves participants fixating on 25 different positions on a screen \cite{kastrati2021eegeyenet}. EEGEyeNet provides both minimally and maximally pre-processed data. We focus on the minimally pre-processed data. This data includes trials from 27 participants and a total of 21464 samples. Following EEGViT~\cite{yang2023vit2eeg}, we split 70\% of these samples into the training set, 15\% into the validation set, and 15\% into the test set.

\subsection{Training}
\label{training}
We train our models on Google Colaboratory with 1 NVIDIA A100 GPU. Table \ref{training setting} shows our training settings. For pre-training, we employ a larger learning rate decay step size and train for more epochs than during fine-tuning. This is because the reconstruction task is more complicated than the downstream gaze estimation task. For fine-tuning, our settings are consistent with EEGViT. The reason is that we use EEGViT model as our MAE encoder, and the goal is to evaluate the encoder's performance increase solely brought by MAE pre-training. This consistent approach ensures that we are making a fair comparison.

\begin{table}
\begin{center}
\caption{Pre-training and fine-tuning settings.}\label{training setting}
\begin{tabular}{|l|l|l|}
\hline
 &  Pre-training & Fine-tuning\\
\hline
optimizer &  Adam~\cite{kingma2014adam} & Adam\\
base learning rate (lr) &  1e-4 & 1e-4\\
batch size & 64 & 64\\
lr decay step size & 10 & 6\\
lr decay factor & 0.1 & 0.1\\
epochs & 30 & 15\\
\hline
\end{tabular}
\end{center}
\end{table}

\section{Results}
We study the effects of masking ratio and decoder architecture, and report the root mean squared error (RMSE) on the test set. The RMSE is in millimeters (mm). See Section \ref{task} for details of the gaze estimation task.

Each pre-training epoch takes approximately 2.4 to 2.6 minutes. A higher masking ratio takes slightly more time. Each fine-tuning epoch takes approximately 2 minutes.

\begin{figure}[htbp]
    \centering
    \subfloat[MLP decoder]{
        \includegraphics[width=0.8\textwidth]{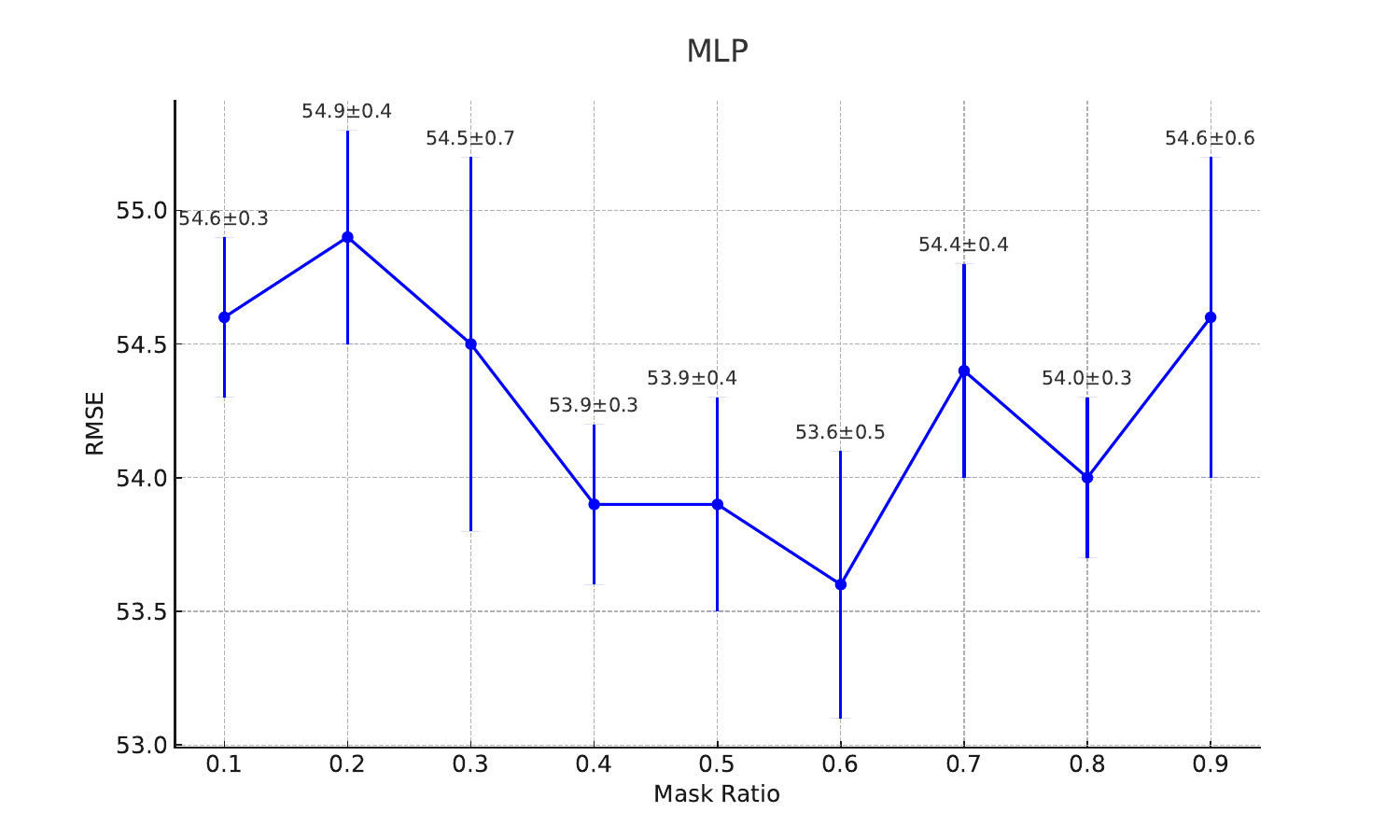}
        \label{fig:sub1}
    }
    \vspace{1mm}
    \subfloat[1 Transformer block decoder]{
        \includegraphics[width=0.8\textwidth]{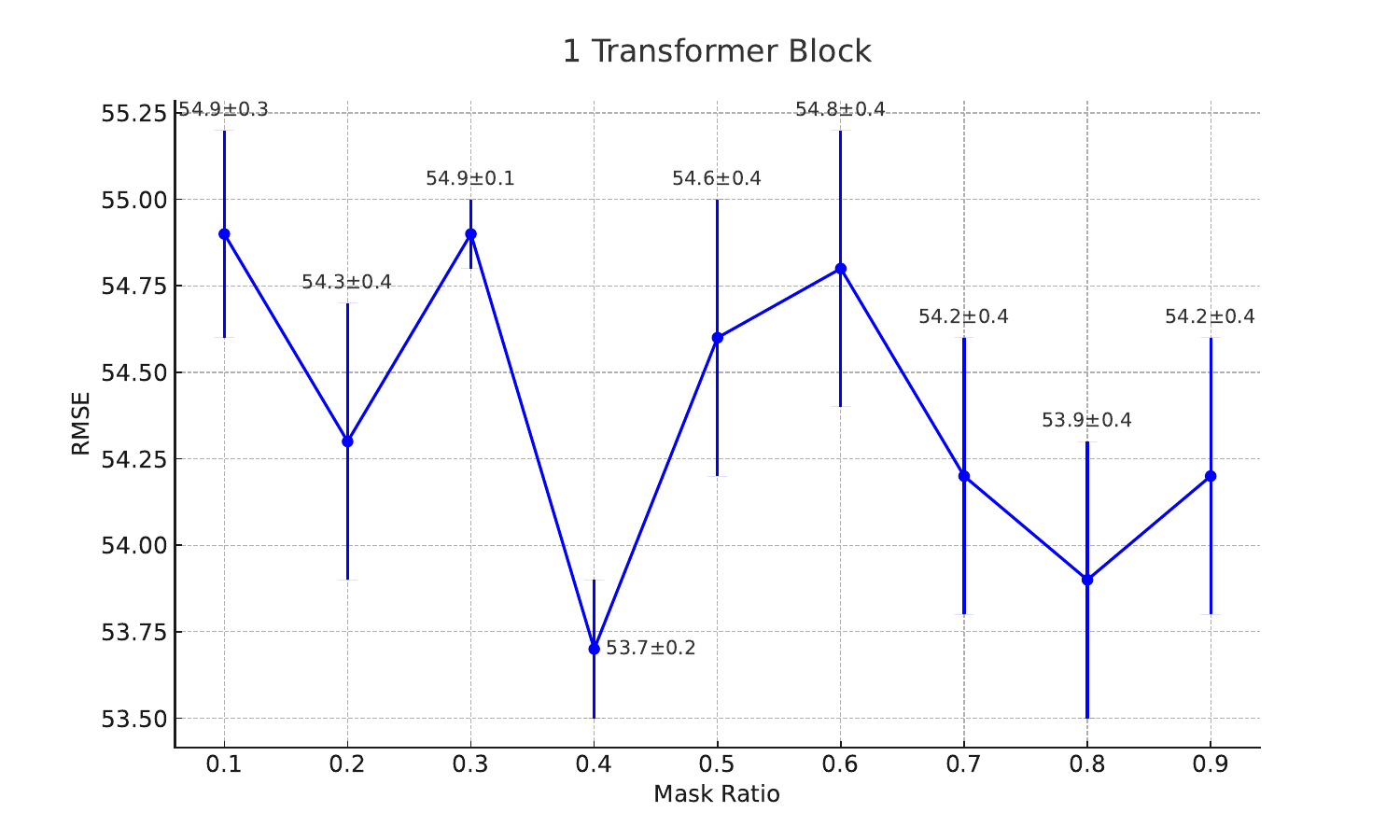}
        \label{fig:sub2}
    }
    \vspace{1mm}
    \subfloat[2 Transformer blocks decoder]{
        \includegraphics[width=0.8\textwidth]{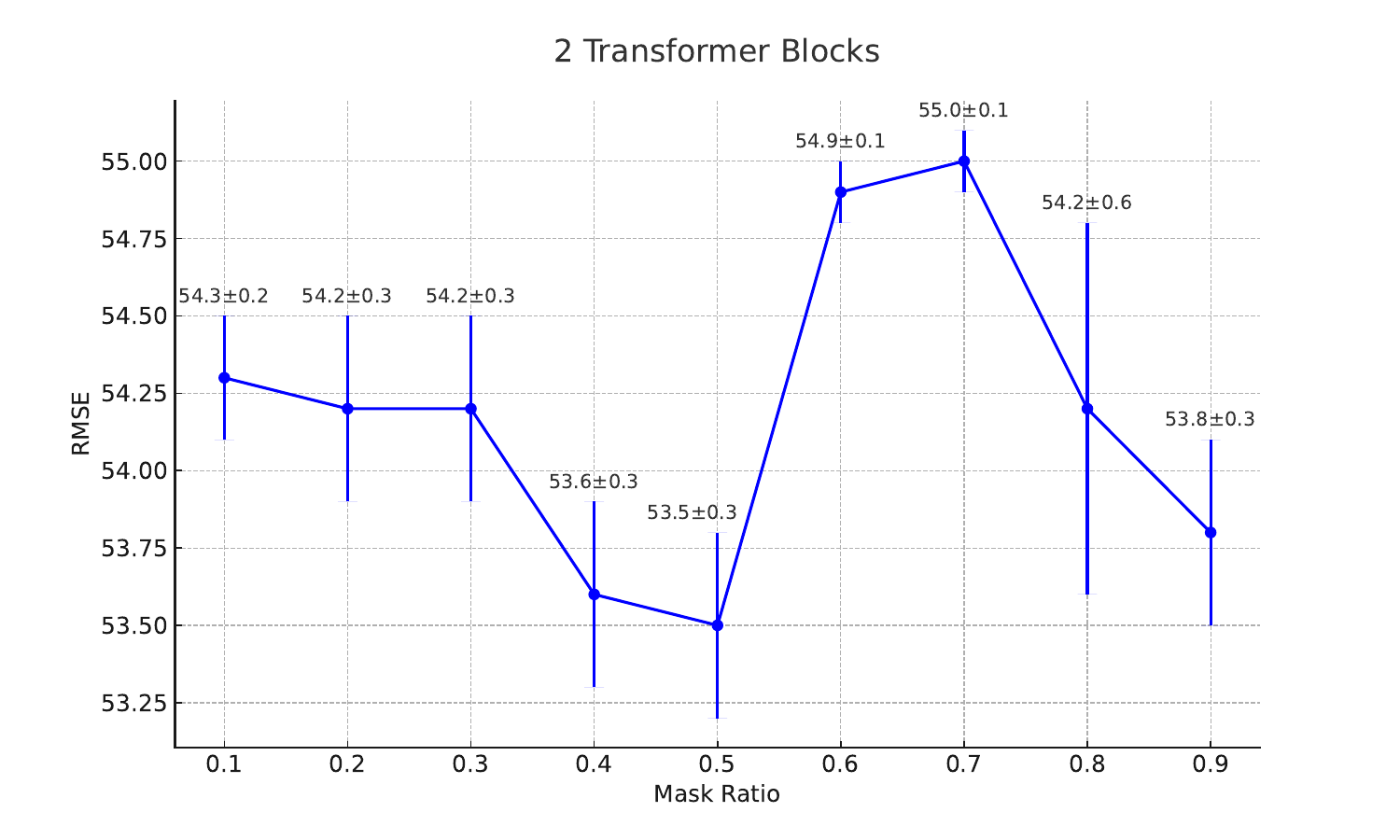}
        \label{fig:sub3}
    }
    \caption{\textbf{Fine-tuning results under different settings.}}
    \label{fig:ratios}
\end{figure}

\begin{figure}[htbp]
    \centering
    \subfloat[MLP decoder]{
        \includegraphics[width=0.8\textwidth]{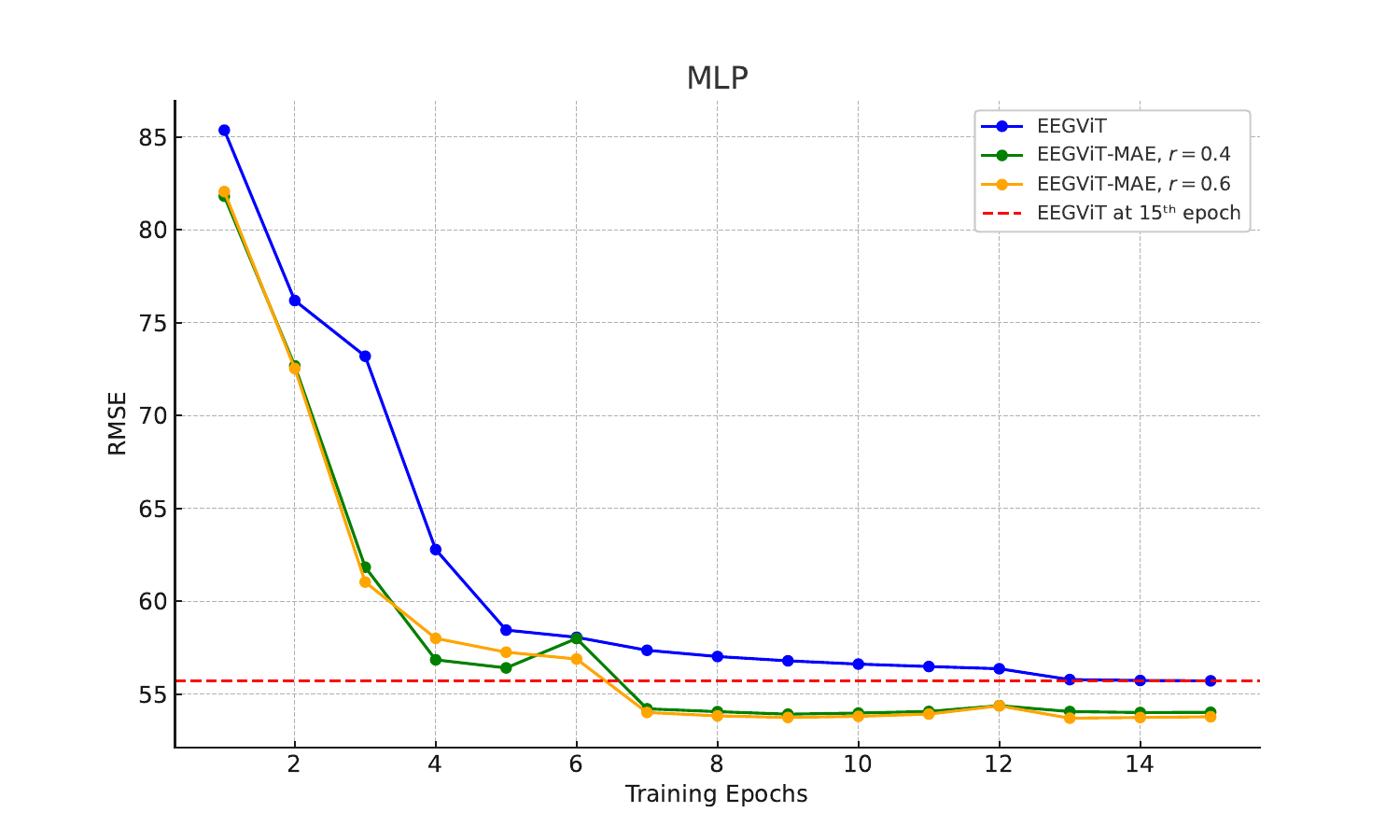}
        \label{fig:sub4}
    }
    \vspace{1mm}
    \subfloat[1 Transformer block decoder]{
        \includegraphics[width=0.8\textwidth]{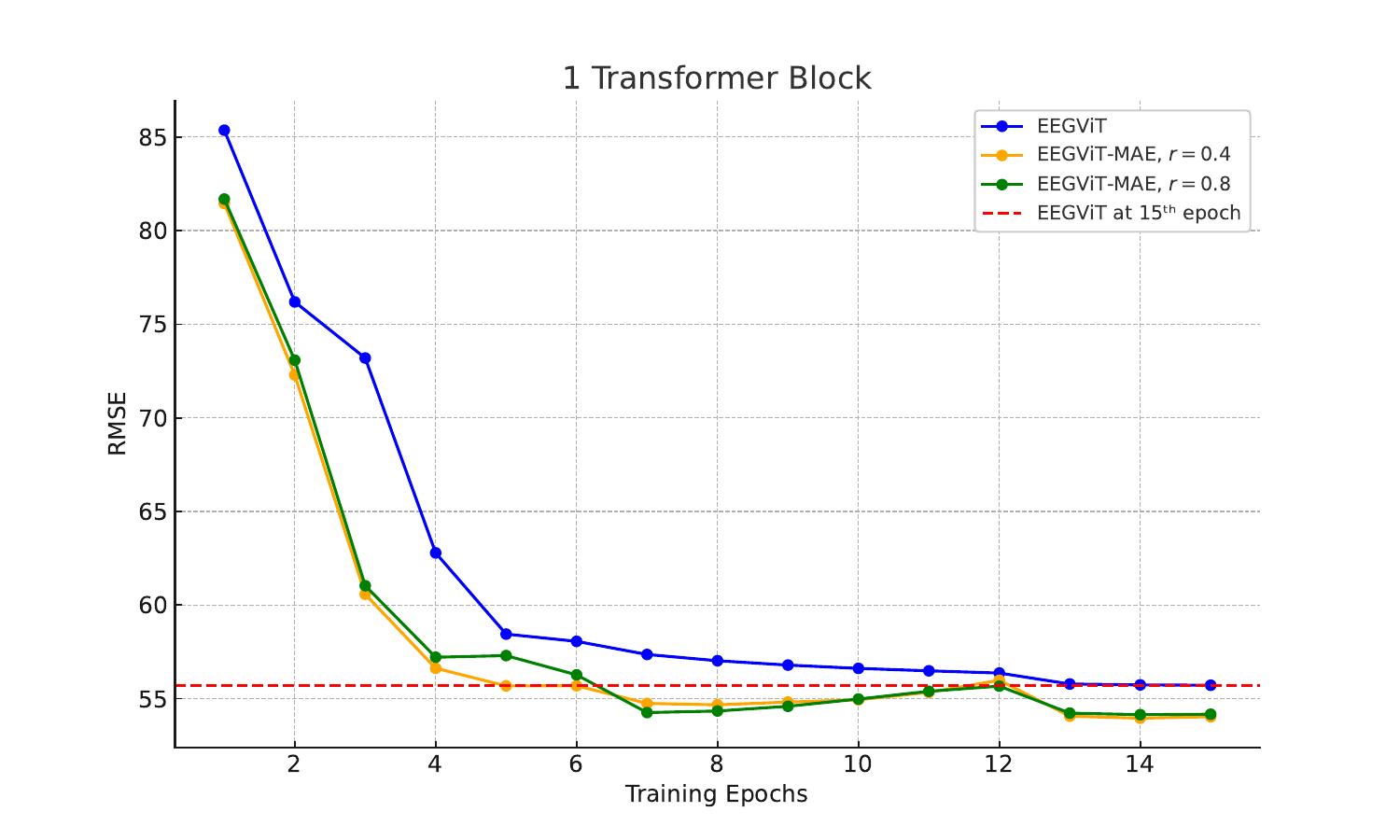}
        \label{fig:sub5}
    }
    \vspace{1mm}
    \subfloat[2 Transformer blocks decoder]{
        \includegraphics[width=0.8\textwidth]{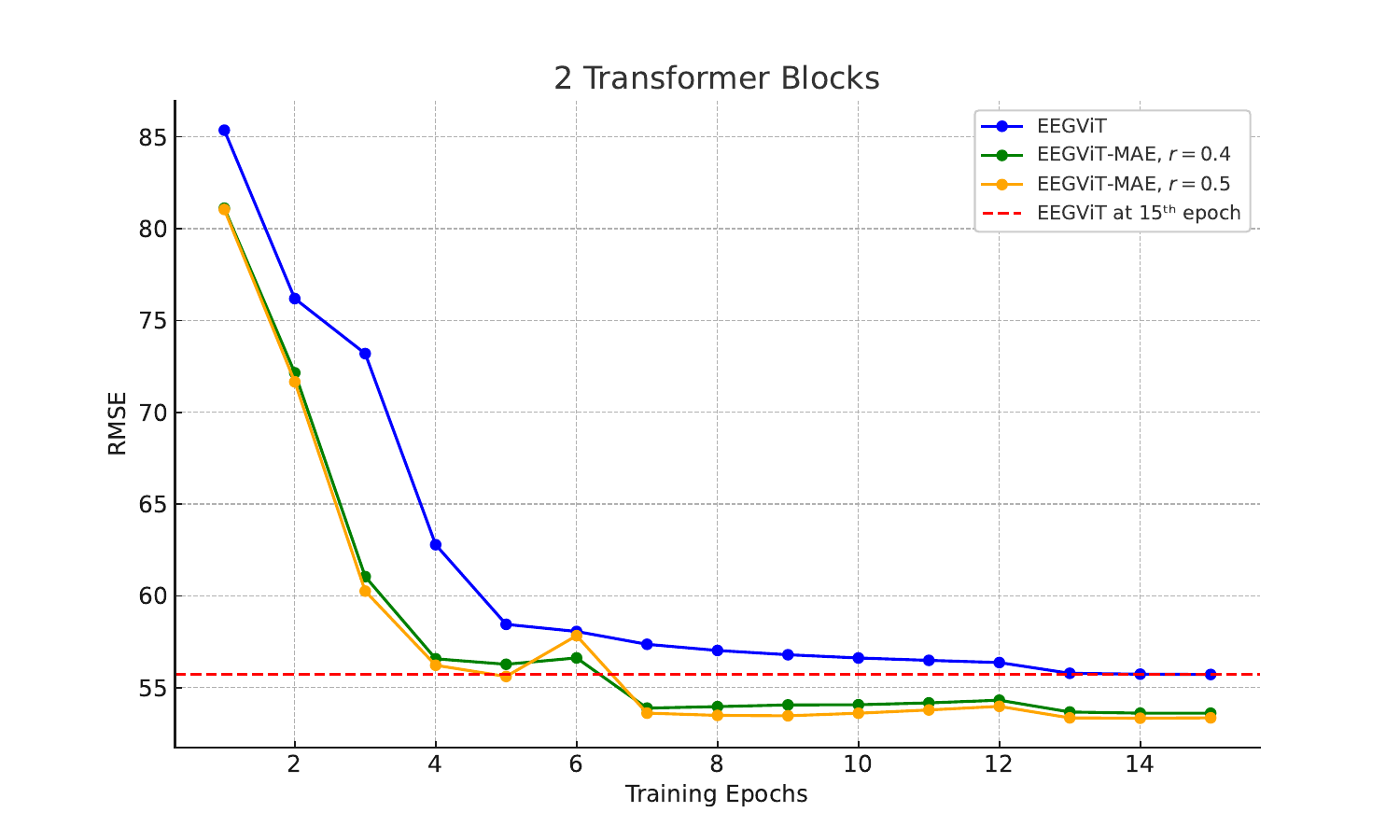}
        \label{fig:sub6}
    }
    \caption{\textbf{Fine-tuning loss curves.} For each decoder setting, top two results among all the masking ratios ($r$) are presented.}
    \label{fig:efficiency}
\end{figure}

\subsection{Encoder's performance}
For MAE pre-training, we experiment with different masking ratios (10\%--90\%). The MAE decoder has 1 or 2 Transformer blocks. In Section \ref{decoder}, we hypothesize that our reconstruction task needs a non-trivial decoder architecture. Here, we also use a simple multilayer perceptron (MLP) decoder as a baseline. Table \ref{results} shows the mean and standard deviation over 5 fine-tuning runs. EEGViT's result is from our experiment\footnote{Here "EEGViT" is equivalent to "EEGViT Pre-trained" in Table 4 of~\cite{yang2023vit2eeg}. This applies to the following mentions as well.}. For each decoder architecture, the best result among all the masking ratios is presented in the table. See Figure \ref{fig:ratios} for the full results.

We find that MAE pre-training reduces the encoder's prediction error without extra hyperparameter tuning. MAE decoder with 2 Transformer blocks achieves the lowest average RMSE. However, the best results of these three decoder architectures are fairly close. From Figure \ref{fig:ratios}, we see that the encoder's variance on the gaze estimation task tends to be lower when pre-trained along with more complex decoders, indicating that non-trivial decoder architectures help stabilize the fine-tuning. We also notice that masking 40\% of the input signal gives relatively good results in all these three decoder settings. We infer that a masking ratio between 40\% and 50\% is the optimal choice for our MAE.

\begin{table}
\begin{center}
\caption{Results from 5 fine-tuning runs.}\label{results}
\begin{tabular}{|l|c|}
\hline
Model & RMSE (mm)\\
\hline
EEGViT &  55.9 ± 0.7\\
EEGViT-MAE, MLP &  53.6 ± 0.5\\
EEGViT-MAE, 1 Transformer Block & 53.7 ± 0.2\\
EEGViT-MAE, 2 Transformer Blocks & \textbf{53.5 ± 0.3}\\
\hline
\end{tabular}
\end{center}
\end{table}

\subsection{Encoder's efficiency}
As discussed in Section \ref{training}, our supervised fine-tuning setting is consistent with EEGViT supervised training. We have shown that EEGViT pre-trained with our MAE achieves better results within the same training epochs. This suggests that it adapts faster to the gaze estimation task after MAE pre-training. Figure \ref{fig:efficiency} shows the fine-tuning loss curves. For each decoder setting, top two results among all the masking ratios are presented.

We find that after MAE pre-training, EEGViT achieves better performance with half the training epochs. For masking ratio $r=0.4$ in the 1 Transformer block setting and $r=0.5$ in the 2 Transformer blocks setting, EEGViT achieves equal performance with 1/3 of training epochs. This demonstrates a significant improvement in learning efficiency. We also observe mild overfitting in EEGViT-MAE models, but it is mitigated in the 2 Transformer blocks setting.

\section{Discussion and Conclusion}
Visual knowledge that is learned from large image datasets like ImageNet can be transferred to the EEG domain, which indicates that these two different signals share some common underlying characteristics. Masked autoencoders (MAEs) are capable of learning useful visual representations. We show that MAEs are effective brain signal learners as well. MAE pre-training is beneficial to downstream tasks in terms of prediction precision and learning efficiency. In this work, we use the EEGViT model as the MAE encoder. However, we expect MAE pre-training to be a generalizable approach to learn EEG signal representations. The encoder model's choice is flexible.In our future work, we plan to explore alternative encoder models beyond EEGViT to evaluate the generalizability of MAE pre-training. Additionally, we aim to extend our experiments to include a wider range of EEG datasets. Furthermore, we intend to investigate other potential deep learning approaches on various datasets for comparative analysis \cite{an2023survey,an2023transfer,jiang2023successfully,lu2023machine,chen2024trialbench,gui2024remote,ma2022traffic,ma2024data,tan2021multivariate,tan2023audio,qiu2023modal,zhang2023trep,zhang2022attention,zhao2024deep}.

Self-supervised pre-training has been widely explored in NLP and computer vision. Similarly, EEG signal research could take this path by building large and diverse EEG datasets to pre-train deep learning models. These pre-trained models can serve as foundation models \cite{bommasani2021opportunities,yang2023foundation,zhou2023comprehensive,li2023multimodal,firoozi2023foundation} for EEG-based applications. They can be fine-tuned on downstream tasks and are expected to obtain superior performance and efficiency compared to models trained solely with supervised learning.

\vspace{10mm}
\begin{credits}
   \textbf{\discintname} The authors have no competing interests to declare that are relevant to the content of this article.
\end{credits}

%
%

\bibliographystyle{splncs04}
\bibliography{mybibliography}

\end{document}